# PA-iMFL: Communication-Efficient Privacy Amplification Method against Data Reconstruction Attack in Improved Multi-Layer Federated Learning

Jianhua Wang, Xiaolin Chang, Jelena Mišić, Vojislav B. Mišić, Zhi Chen, and Junchao Fan

*Abstract*—Recently, big data has seen explosive growth in the Internet of Things (IoT). Multi-layer FL (MFL) based on cloud-edge-end architecture can promote model training efficiency and model accuracy while preserving IoT data privacy. This paper considers an improved MFL, where edge layer devices own private data and can join the training process. iMFL can improve edge resource utilization and also alleviate the strict requirement of end devices, but suffers from the issues of Data Reconstruction Attack (DRA) and unacceptable communication overhead.

This paper aims to address these issues with iMFL. We propose a *P*rivacy *A*mplification scheme on *iMFL* (PA-iMFL). Differing from standard MFL, we design privacy operations in end and edge devices after local training, including three sequential components, local differential privacy with Laplace mechanism, privacy amplification subsample, and gradient sign reset. Benefitting from privacy operations, PA-iMFL reduces communication overhead and achieves privacy-preserving. Extensive results demonstrate that against State-Of-The-Art (SOTA) DRAs, PA-iMFL can effectively mitigate private data leakage and reach the same level of protection capability as the SOTA defense model. Moreover, due to adopting privacy operations in edge devices, PA-iMFL promotes up to 2.8 × communication efficiency than the SOTA compression method without compromising model accuracy.

## I. INTRODUCTION

Machine Learning (ML) has become an indispensable composition of enhancing industrial productivity and improving daily life, such as medical diagnosis [1], and network traffic analysis [2]. Abundant training data ensures the effectiveness of an ML model. Hence, a huge number of Internet of Things (IoT) devices are set for collecting data, such as vehicle driving data, transportation traffic, and charging records of electric vehicles.

However, personal data leakage accidents frequently occur, and it becomes troublesome to exploit private data without privacy-preserving. Moreover, data protection laws have been published by governments to constrain the private usage of data, such as the General Data Protection Regulation (GDPR) [3].

**Standard FL**. Driven by the increasingly privacy-preserving demands, Federated Learning (FL) [4], proposed by Google, as a burgeoning distributed ML paradigm for preventing private information leakage, has gained tremendous concentrations. Specifically, in a standard FL system with multi-party, participants, such as the enterprise, financial institution, and vehicle charging station, train their private dataset locally, and then transmit the model update to an honest-but-curious third-party parameter server for model aggregation. Thus, FL provides a paradigm of training collaboratively without data sharing.

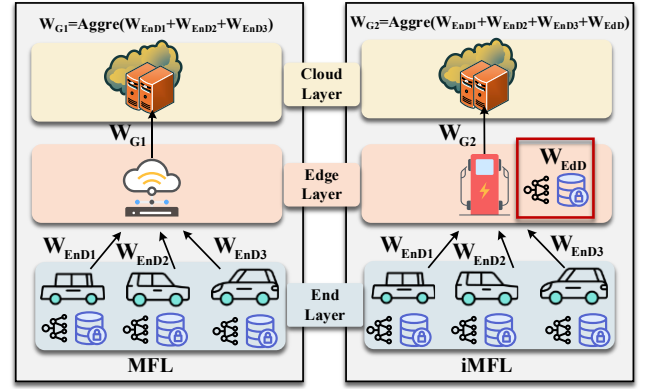

**Fig.1**. The illustration of the MFL and iMFL.

**Traditional Multi-layer FL**. Except for the standard FL with a two-layer structure, studies have recently explored the Multi-layer FL (MFL) based on cloud-edge-end three-layer architecture [5]–[7], illustrated in **Fig.1**. Each edge device only aggregates local model updates and sends it to cloud layer. The MFL can greatly reduce communication overhead, exploit the high bandwidth with short physical distances [6], and promote training efficiency and model accuracy [8]. Thus, it is suitable for a wireless network environment. More importantly, the cloud server, in MFL, avoids network congestion while standard FL does not due to the large number of participants accessing simultaneously [6]. However, the MFL at least has two issues.

(i) Low utilization of edge computation resources. The MFL assumes that the edge layer is only responsible for aggregating the model received from the end layer and uploading the real-time update to the cloud layer. Compared to mobile devices in the end layer, the edge layer devices are located in a fixed position, such as a service base station [6] and edge server [5], with more computation resources than the end layer. However, the edge layer devices do not utilize their computation resources efficiently.

(ii) Strict requirement of the number of end devices. Actually, the MFL requires no less than one participant in the training process, otherwise, the training process will be suspended. This assumption is not suitable for mobile end devices, such as unmanned aerial vehicles, intelligent vehicles, and laptops, which join the training process dynamically and dominate limited resources.

**Motivation and improved MFL**. To address the above-mentioned issues of traditional MFL, we extend the MFL and then propose an improved MFL (iMFL) by adding the resource-constrained edge layer devices into the model training task (see **Fig.1**). In other words, Edge devices, which own their private datasets, not only train their local model but also are responsible for aggregating the local model update received from end devices. Under these circumstances, the edge layer maximizes the utilization of its own resources, and will not rely on the number of end devices in training.

However, iMFL still at least has the following two challenges to be addressed.

**Challenge 1 Data Reconstruction Attack (DRA) in privacy-preserving**. Extensive efforts have been conducted to solve the potential problems of FL, for example, preventing model poison attack [9]–[11], membership inference attack [12], and free-rider attack [13]. As an emerging privacy leakage attack paradigm, DRA is a more serious threat [14], in which the adversary reconstructs accurate original data from global shared gradients in FL [15]–[18]. Especially, recent studies indicate that advanced DRA methods recover the high fidelity original images against DRA defense models [17]–[19], such as gradient compressions [20], and perturbed representations [21].

**Challenge 2 Communication Overhead in downlink and uplink between end and edge devices**. Recently, with the flourishing demands of Deep Learning (DL), the DL model, with billions of parameters, is carried out by end and edge devices. However, due to the mobility of end devices, parameters are uplinked or downlinked from end to edge layer in a limited wireless communication network [22], which raises heavy communication overheads and constitutes the major bottleneck of iMFL [23].

**Prior Works.** Concentrating on the aforementioned challenges, past works provide the solutions. For **Challenge 1**, there exist mainstream defense methods, namely, data augmentation-based [14], Differential Privacy (DP)-based [15], [21], cryptology-based [24], gradient compression-based [25]. However, data augmentation-based and cryptology-based methods deeply rely on abundant computation resources and the perturbation-based methods fail to reveal substantive defending reasons for DRA, which leads the large noise along with low model accuracy [18]. For **Challenge 2**, existing studies focus on gradient compression. However, compared to the sample-based method, the prune-based, quantization-based, knowledge distillation-based, and low-rank factorization-based methods are unbalancing between communication efficiency and model accuracy [26].

**Intuition**. Sampling, as a type of gradient compression method, bridges the privacy-preserving and communication efficiency [27]–[30]. However, existing State-Of-The-Art (SOTA) sample-based work lost effectiveness when facing the DRA [17]. Moreover, some sample methods need extra computation due to the use of Secure Multi-party Computation (SMC) [30]. Actually, privacy amplification by subsample is a crucial primitive for designing DP mechanisms [31]–[33]. Thus, we exploit this principle to defend DRA and reduce the communication overhead in iMFL.

**Our Work**. In this paper, we propose a *P*rivacy *A*mplification scheme in *i*mproved *MFL* (PA-iMFL) to address the two aforementioned challenges in iMFL (illustrated in **Fig.4**). We divide the PA-iMFL into two parts, respectively end layer process (steps 1-6 in end layer) and edge layer process (steps 1-5 in edge layer). Unlikely the standard local training process, PA-iMFL conducts the privacy operations (red steps 2-4 in the end layer, and red step 3 in the edge layer of **Fig.4**), containing three sequential components: (red step 2) Local DP with Laplace mechanism, (red step 3) Privacy Amplification Subsample (PAS), and (red step 4) Gradient Sign Reset (GSR).

Based on LDP with Laplace mechanism, we design PAS and GSR for amplifying the capability of privacy-preserving and achieving succinct transmission. Specifically, PAS utilizes the unbiased subsample by selecting top-k rescaled gradients to decrease the communication overhead remarkably. In GSR, we reset the gradient sign to confuse the adversary in DRA. Moreover, except for past works with sampling in the end layer, we extend the subsample to the edge layer, which means that the edge devices not only receive the subsample local update from the end layer but also broadcast the subsample global update to the end layer. Extensive evaluation results demonstrate that PA-iMFL not only protects the private dataset against SOTA DRA [16]–[18] but also reduces the communication overhead without compromising model accuracy.

**Contributions**. We summarize the unique features of PA-iMFL as follows.

- Feature 1. iMFL is a re-defined three-level FL. To the best of our knowledge, we are the first to define the iMFL framework, where edge devices take part in the FL training process as well. Edge and end devices are resource-constrained devices.
- Feature 2. PA-iMFL is privacy-preserving against DRA. We theoretically prove the privacy guarantee and present the LDP condition of PA-iMFL. Differing from existing privacy amplification works, PA-iMFL adds a GSR component against sample-oriented DRA. Extensive evaluation results demonstrate that against SOTA DRA, the adversary in PA-iMFL hardly recovers the private image of PA-iMFL participants.
- Feature 3. PA-iMFL is communication-efficient. PA-iMFL achieves bi-directional gradient compression in the end layer and edge layer under our iMFL architecture. The result reveals that PA-iMFL reaches up to 3.8× efficiency than the SOTA gradient compression method with a subsample ratio of 0.07, and up to 2.8× efficiency promotion without compromising model accuracy.

The rest of our paper is set up as follows. Section II presents the preliminary of FL and DP. Section III gives the problem statement of this paper, including scene overview, threat model, privacy problem formulation, design goals, and our solution and optimization objective. Section IV and Section V provide the scheme description and main evaluation results, respectively. Section VI discusses related work. In the end, we conclude this paper in Section VII.

## II. PRELIMINARY

This section presents the preliminary of FL and DP.

### A. Federated Learning

A standard FL system usually includes a set of distributed participants and a single cloud central server. Participants collaborate to train an optimal global model without transmitting their private datasets. Eq. (1) gives the overall optimization objective of the simplified FL process.

$$\min_{W \subseteq \mathbb{R}^d} \mathcal{L}(W_G) := \sum_{i \in [m]} \frac{P_i}{\sum_{i \in [m]} P_i} l(W_i) \quad (1)$$

Here, $W_G$ denotes the global model and $W_{\ell_i}$ is the local model of the participant $i$. $P_i$ denotes the personalized weight of the participant $i$, while $\mathcal{L}(\cdot)$ and $l(\cdot)$ are the loss function for the FL system and participants, respectively.

*B. Differential Privacy*

Differential Privacy (DP), first proposed by Dwork *et al.* [34], is widely used in data analytics and plays a crucial role in privacy-preserving FL. Now, we give the definition of DP in Eq. (2) and Local DP in Eq. (3).

**Definition 2.1.** $(\epsilon, \delta)$-**Differential Privacy.** *A randomized mechanism $\mathcal{M}$ satisfies $(\epsilon, \delta)$-DP w.r.t. $\epsilon, \delta \geq 0$ and neighboring dataset $\mathcal{D}$ and $\mathcal{D}'$ iff the output of $\mathcal{M}$ belonging to $\mathcal{O} \subseteq \text{Range}(\mathcal{M})$, and*

$$\Pr[\mathcal{M}(\mathcal{D}) \in \mathcal{O}] \leq e^\epsilon \Pr[\mathcal{M}(\mathcal{D}') \in \mathcal{O}] + \delta. \quad (2)$$

Notably, if $\delta \in [0,1)$ satisfies $\delta = 0$, the mechanism $\mathcal{M}$ is $\epsilon$-DP, which is pure without relaxation probability [35]. Here, $\text{Range}(\mathcal{M})$ denotes the set of all probable outputs $\mathcal{M}$.

The standard interpretation of the definition of $(\epsilon, \delta)$-DP is with the probability $1 - \delta$, namely, the probability ratio of the output of mechanism $\mathcal{M}$ under two neighboring datasets $\mathcal{D}$ and $\mathcal{D}'$ is less than $e^\epsilon$. It demonstrates that little $\epsilon$ and $\delta$ mean less probability of inferring private datasets, and privacy is guaranteed.

**Definition 2.2.** **Local Differential Privacy (LDP).** *A randomized mechanism $\mathcal{M}$ satisfies $(\epsilon, \delta)$-LDP w.r.t. neighboring relation private data $x$ and $x'$ of device client $D$, iff the output of $\mathcal{M}$ belonging to $\mathcal{O} \subseteq \text{Range}(\mathcal{S})$, and*

$$\Pr[\text{Perc}_\mathcal{M}(x) \in \mathcal{O}] \leq e^\epsilon \Pr[\text{Perc}_\mathcal{M}(x') \in \mathcal{O}] + \delta. \quad (3)$$

Here, $\text{Perc}_\mathcal{M}(\cdot)$ denotes the central server perception of the device client's information. Notably, the server usually is honest-but-curious, leading to inferring private information.

## III. PROBLEM STATEMENT

This section first states the overview of iMFL. Then, the threat model is presented. After that, problem formulation is defined. Moreover, the design goals of the defense model are given, and at last, we present our solution and optimization objective.

*A. iMFL Overview*

**Fig.2** illustrates the overview of the iMFL. Same as traditional MFL, iMFL also includes three layers: the end layer, edge layer, and cloud layer. End and edge are commonly resource-constrained devices.

- **End Layer.** The end layer contains end devices such as smart electric vehicles, which are the principal FL participants and global model trainers in iMFL with local privacy datasets.
- **Edge Layer.** The edge layer includes various edge devices like intelligent roadside units, transportation surveillance cameras, intelligent charging stations, parking lots, etc. Similar to end devices, edge devices have private datasets. For example, the intelligent charging station has its privacy service data or charging work orders. In other words, an edge device, as an honest-but-curious third party, is not only responsible for model aggregation from the end layer but also for training the global model using its dataset to improve the model.
- **Cloud Layer**. The cloud layer aggregates the global update received from the edge layer. Notably, the cloud layer includes the cloud computing center, which is the summit manager of iMFL and dominates abundant computing, communication, and storage resources.

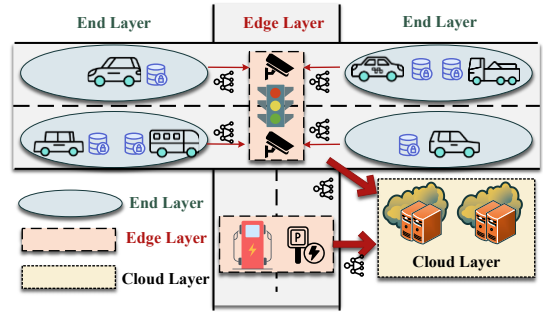

**Fig.2**. The overview of the iMFL.

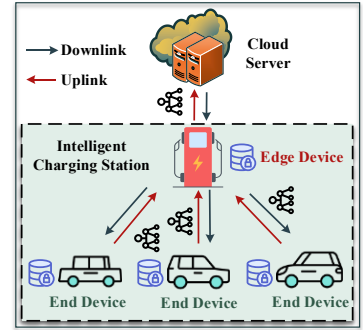

**Fig.3**. The iMFL workflow in the Intelligent Charging Station scene.

We take the intelligent charging station scene as an example in **Fig.3**. The red arrow denotes the model uplink direction while the blackish-green arrow denotes the model downlink direction. Intelligent vehicles are charging in an Intelligent Charging Station. End and edge devices train the broadcasted global model using their private datasets. After training locally, the edge device aggregates the received local updates and its local update using an aggregation algorithm. Then, the edge device transmits the aggregated global update to the cloud layer for final aggregating.

*B. Threat Model*

In this paper, we consider that each iMFL participant can be an honest-but-curious adversary, such as end devices, edge devices, and cloud servers. They obey the iMFL workflow, but they are curious to obtain the local model. Specifically, the adversary we define possesses the following characteristics:

- **Transmit update honestly.** The adversary in the edge layer transfers the correct local updates honestly rather than modifying them. Intuitively, an adversary wants to maintain the performance of the iMFL to obtain the maximum benefit, including a well-trained global model, and long-time camouflage.
- **Eavesdrop update promptly.** The adversary in the end, edge, and cloud layer, eavesdrops and records the update immediately. Actually, for training demands, each iMFL participant will record the update. However, it may contain abundant information and the private dataset may be leaked by DRA.
- **Possess knowledge expertly.** Generally, the adversary only possesses the model information of some participants, called auxiliary-free [36] DRA. However, we assume the adversary has expert knowledge, called auxiliary-based [36] DRA. In other words, the adversary not only holds the model information but also owns the pre-trained model by additional auxiliary datasets, such as the generative adversarial network. This kind of adversary has a higher attack capability.

*C. Privacy Problem Formulation*

We only consider the iMFL including end devices (*EnD*) and edge devices (*EdD*) $\mathbb{P} = \{EnD_1, EnD_2, ..., EnD_i, ..., EnD_m, EdD\}$. In a small scene like the Intelligent Charging Station in **Fig.3**, there are several EnDs and one EdD. Each participant owns their private dataset $\mathcal{D} = \{\mathcal{D}_{EnD1}, ..., \mathcal{D}_{EnDi}, ..., \mathcal{D}_{EdD}\}$, where $\mathcal{D}_i := \{(x_{i,j}, y_{i,j}) | j \in (1, n_i)\}$. Here, $n_i$ denotes the data number of the dataset $\mathcal{D}_i$, and $(x_{i,j}, y_{i,j})$ is a pair of data and a responding label. We denote $\mathcal{L}(W_G; (X,Y)): W \to \mathbb{R}$ as the loss function of the global model and $W \subset \mathbb{R}^d$ is the weight parameter space. Similarly, $l(W^r_{EnD_i}; (x_{i,j}, y_{i,j}))$ denotes the local loss function of $EnD_i$, and $l(W_{EdD}; (x,y))$ denotes the local loss function of $EdD$. Hence, the optimization objective of the FL process with the FedAvg aggregation strategy [4] is stated as in Eq. (4), and the optimal parameter $W^*$ satisfies Eq. (5).

$$\min_{W \subseteq \mathbb{R}^d} \mathcal{L}(W_G; (X,Y)) := \frac{1}{m+1} \left( \sum_{i=1}^{m} l(W_{EnD_i}; (x_{i,j}, y_{i,j})) + l(W_{EdD}; (x,y)) \right) \quad (4)$$

$$W^* = \arg\min \mathcal{L}(W_G; (X,Y)) \quad (5)$$

However, the adversary $\mathcal{A}$ eavesdrops on the model information to reconstruct the private dataset. For example, assume that the edge device $\mathcal{A}_{EdD}$ can obtain the $W_{EnD_i}$ of each $EnD_i$. The DRA can be formulated as in Eq. (6).

$$(x^*, y^*) = \arg\min_{x', y'} \left\| \nabla l(W_{EnD_i}; (x_{i,j}, y_{i,j})) - \nabla l(W_{\mathcal{A}_{EdD}}; (x', y')) \right\|^2 \quad (6)$$

where $(x^*, y^*)$ is the optimal reconstructed data and responding label. $(x', y')$ denotes the dummy data, which may be initialized by Gaussian distribution [15] or other pre-trained generative networks [18].

*D. Design Goals*

In this paper, we concentrate on iMFL in resource-constrained devices, such as intelligent vehicles. On the one hand, we aim to achieve computation-friendly and communication-efficient iMFL. On the other hand, to protect the privacy dataset of participants, our iMFL needs to resist the state-of-the-art DRA. Specifically, our scheme PA-iMFL needs to meet the following design goals.

- **Privacy Preservation (Goal 1).** The iMFL provides a data-protecting paradigm. However, DRA infers the private dataset using transmitting gradients. Our scheme must protect the privacy of participants against DRA in the iMFL training process.
- **Communication Efficiency (Goal 2).** Since numerous participants are resource-constrained, our scheme cannot provoke extra computation costs and communication overhead. The uplink and downlink between end devices and edge devices are succinct.
- **Performance Guarantee (Goal 3).** The adversary obeys the honest-but-carious protocol claimed in Section III.B, which implies the adversary cannot produce a negative impact on the iMFL. Hence, the succinct communication we achieve is performance-guarantee.

*E. Our Solution and Optimization Objective*

Different from the general FL training processing with abundant communication and computation resources, iMFL needs to satisfy the goals aforementioned using the PA-iMFL scheme (detailed descriptions in Section IV.A).

First of all, to satisfy **Goal 1**, the PA-iMFL should protect the private datasets of end and edge devices against DRA.

Then, as illustrated in **Fig.4**, the edge layer is in charge of the local training, model aggregation, update uploading, and initialized/update broadcasting. Hence, unlike the existing methods, we extend the subsample [23], [27], [29] to bi-directional subsample in PA-iMFL to achieve **Goal 2**. In other words, we conduct subsamples twice in the edge layer, namely the first subsampled in uploading the local update and aggregating the received update to the cloud layer, and the second subsampled in broadcasting the aggregated global model to the end layer.

Finally, to guarantee the accurate performance of the whole system and satisfy **Goal 3**, in the end and edge layers, we utilize the local momentum technique in PA-iMFL to compensate for the lost information of the local gradient because of privacy operations.

In this section, we will present the optimization problem of our PA-iMFL scheme. Notably, we utilize the privacy operation sets $\mathcal{P} := \{P_1, P_2, P_3\} = \{\text{Lap}(\frac{\Delta f}{\epsilon}), PAS(\nabla W, \gamma), GSR(\nabla W, t)\}$ in the end and edge layers. $\text{Lap}(\cdot)$ denotes the LDP with Laplace mechanism, namely $P_1$. $PAS(\cdot)$ is the subsample mechanism, namely $P_2$, and $GSR(\cdot)$ represents the gradient sign reset mechanism, namely $P_3$. Besides, $\Delta f$ is the sensitivity of each query. We utilize $\nabla W$ to represent the gradient of the local model. $\epsilon_1$ and $\epsilon_2$ are the privacy budget, $\delta$ is relaxation probability, and $\gamma$ denotes the subsample ratio. In $GSR(\cdot)$, $t$ is the interval rounds of GSR. Assuming that $d$ is the length of the whole gradient, we can find the optimization objective of PA-iMFL as follows in Eq. (7).

$$\min_{W,\epsilon,\delta} \mathcal{L}(W_G;(X,Y)) := \frac{1}{m+1}\left(\sum_{i=1}^{m} l(W_{EnD_i};(x_{i,j},y_{i,j});\mathcal{P}_{EnD}) + l(W_{EdD};(x,y);\mathcal{P}_{EdD})\right)$$

$$s.t. \ \delta \ll \frac{1}{d} \text{ and } \delta \in [0,1], \ t \geq 0, \ \gamma \in (0,1] \quad (7)$$

$$\max \left\| \nabla l(W_{EnD_i};(x_{i,j},y_{i,j})) - \nabla l(W_{\mathcal{A}_{EdD}};(x',y')) \right\|^2$$

with $y_{i,j} = y'$

Notably, we minimize $\mathcal{L}(\bullet)$ subject to four constraint conditions. According to [5], [8]-[9], the LDP relaxation probability should be $\delta \ll 1/d$ [38] to satisfy the privacy-preserving requirements. Moreover, we have to maximize the divergence between the original input and reconstructed data. Notably, numerous studies focus [3], [10]-[11] on the label inferring to obtain the true label of target data for better-reconstructed dummy data. However, in this paper, we suppose that the adversary $\mathcal{A}$ has inferred the true label $y'$ of target data. This means that besides the expert knowledge, the adversary $\mathcal{A}$ has unrivaled attack capability. Notably, we define $\mathcal{G} := \nabla W$, and in the latter descriptions, we use $\mathcal{G}$ instead of $\nabla W$.

## IV. PA-iMFL Scheme Description

This section presents a detailed description of the PA-iMFL scheme firstly. Then, we give privacy operations, namely the LDP with Laplace mechanism, privacy amplification subsample mechanism, and gradient sign reset mechanism. At last, the privacy analysis of PA-iMFL is discussed.

### A. Description of PA-iMFL

As illustrated in **Fig.4**, we present the training process of PA-iMFL in three layers, including the end layer, edge layer, and cloud layer. Notably, end devices are the main participants in iMFL training while edge devices not only are responsible for aggregating the local update from the end layer but train the global model using their datasets.

*1) End Layer Steps*

Steps 1, and 5 are the same as in the standard FL workflow in [4], step 6 is spired by [39], and steps 2,3,4 are proposed in this paper. For easy understanding, we detail each step in the following.

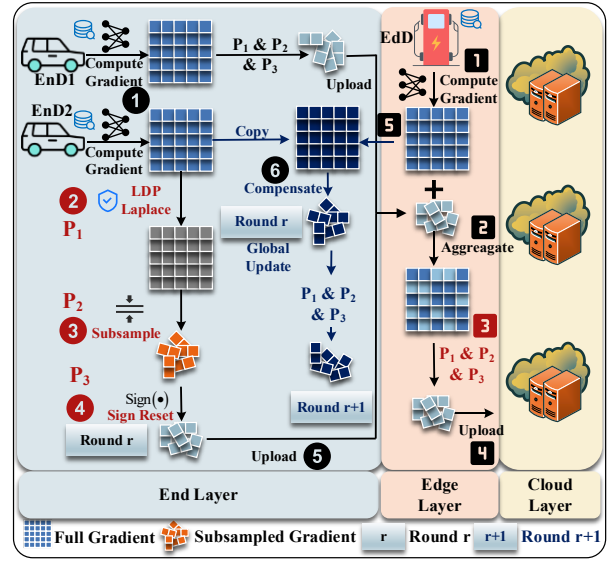

**Fig.4.** Illustration of the PA-iMFL scheme. Red steps 2, 3, and 4 in the end layer and red step 3 in the edge layer are proposed in this paper. Steps 1 and 5 in the end layer and steps 1, 2, and 4 in the edge layer, follow the standard FL workflow. Step 6 in the end layer and step 5 in the edge layer are spired by [39].

**Step ❶: Gradient Compute and Copy**. In the training round $r=0$, the $EnD_i$ in $\mathbb{P}$ receives the whole global model $W_G^0$ from the edge layer and transfers the corresponding local model $W_{EnD_i}^0$. Otherwise, $EnD_i$ obtains the global update $\Delta \mathcal{G}_G^{r-1}$ in round $r-1$ and computes the local model by SGD definition $W_{EnD_i}^{r-1} \leftarrow W_{EnD_i}^{r-2} - \eta(\Delta \mathcal{G}_G^{r-1} + \mathcal{G}_G^{r-2})$. Then, $EnD_i$ trains the local model $W_{EnD_i}^{r-1}$ using its private dataset and calculates the local loss $l(W_{EnD_i}^{r-1})$ by validation process. Moreover, $EnD_i$ computes the local gradient $\mathcal{G}_{EnD_i}^r$ of the round $r$. At last, $EnD_i$ obtains the local gradient update $\Delta \mathcal{G}_{EnD_i}^r \leftarrow \mathcal{G}_{EnD_i}^r - \Delta \mathcal{G}_G^{r-1}$. There is the traditional gradient computing process. In PA-iMFL, spired by [39], we have the extra process in that we copy the local gradient update and store it on the local side $EnD_i$ to compensate for the lost gradient information due to the privacy operation $\mathcal{P}$, namely $R_{EnD_i}^r \leftarrow \Delta \mathcal{G}_{EnD_i}^r$.

**Step ❷: Laplace LDP Conduct (P₁)**. Before transferring, the local update $\Delta \mathcal{G}_{EnD_i}^r$, $EnD_i$ utilizes the LDP with Laplace mechanism to add minuscule perturbation $\Delta \mathcal{G}_{EnD_i}^r \leftarrow \Delta \mathcal{G}_{EnD_i}^r + \varepsilon$, where $\varepsilon \sim \text{Lap}(\frac{\Delta f}{\epsilon})$.

**Step ❸: Privacy Amplification Subsample (P₂)**. After the **P₁** operation, we utilize the subsample mechanism for privacy amplification. Roughly speaking, we conduct the subsample with subsample ratio $\gamma$ and compute the local update by $\Delta \mathcal{G}_{EnD_i}^r \leftarrow PAS(\Delta \mathcal{G}_{EnD_i}^r, \gamma)$. The details of **P₂** are described in Section IV.B.

**Algorithm 1 PA-iMFL in End Layer**

**Input:** training round $R$, end device client $EnD$, number of $EnD$ $m$, **P₁** privacy budget $\epsilon$, subsample ratio $\gamma$, interval rounds $t$, learning rate $\eta$, momentum coefficient $\beta$

**Initialize:** global model $W_G^0$

1 ############### End Layer Process ###############
2 **For** $r = 0,1,2,...,R$ **do**
3   # **STEP ❶**: Gradient Compute and Copy
4   **If** $r == 0$ **then**:
5      $W_{EnD_i}^0 \leftarrow W_G^0$
6   **Else** round $r-1$:
7      Receive global update $\Delta \mathcal{G}_G^{r-1}$
8      $W_{EnD_i}^{r-1} \leftarrow W_{EnD_i}^{r-2} - \eta \left( \Delta \mathcal{G}_G^{r-1} + \mathcal{G}_G^{r-2} \right)$
9   **End if**
10   **For** $EnD_i (i=1,...,m)$ **in parallel do**
11      $\mathcal{G}_{EnD_i}^r \leftarrow \nabla l \left( W_{EnD_i}^{r-1} \right)$
12      $\Delta \mathcal{G}_{EnD_i}^r \leftarrow \mathcal{G}_{EnD_i}^r - \Delta \mathcal{G}_G^{r-1}$
13      Record $R_{EnD_i}^r \leftarrow \Delta \mathcal{G}_{EnD_i}^r$
14      # **STEP ❷**: Laplace LDP Conduct (**P₁**)
15      $\Delta \mathcal{G}_{EnD_i}^r \leftarrow \Delta \mathcal{G}_{EnD_i}^r + \varepsilon$, $\varepsilon \sim \text{Lap}(\frac{\Delta f}{\epsilon})$
16      # **STEP ❸**: Privacy Amplification Subsample (**P₂**)
17      $\Delta \mathcal{G}_{EnD_i}^r \leftarrow PAS(\Delta \mathcal{G}_{EnD_i}^r, \gamma)$
18      # **STEP ❹**: Gradient Sign Reset (**P₃**)
19      $\Delta \mathcal{G}_{EnD_i}^r \leftarrow GSR(\Delta \mathcal{G}_{EnD_i}^r, t)$
20      # **STEP ❺**: Gradient Update Upload
21      Upload $\Delta \mathcal{G}_{EnD_i}^r$ to the $EdD$
22      **If** $r = r+1$ **then**:
23        # **STEP ❻**: Gradient Compensate
24        $\Delta \mathcal{G}_{EnD_i}^{r+1} \leftarrow \beta R_{EnD_i}^r + (1-\beta) \Delta \mathcal{G}_G^r$, $\beta \in [0,1)$
25      **End if**
26   **End for**
27 **End for**

**Step ❹: Gradient Sign Reset (P₃)**. In **P₃**, we randomly change the gradient sign with the interval rounds $t$ to confuse the adversary $\mathcal{A}$ using $\Delta \mathcal{G}_{EnD_i}^r \leftarrow GSR(\Delta \mathcal{G}_{EnD_i}^r, t)$. Similarly, the detail of **P₃** is described in Section IV.B.

**Step ❺: Gradient Update Upload**. $EnD_i$ uploads the local update $\Delta \mathcal{G}_{EnD_i}^r$ to the $EnD$. Up to now, the training process has finished in the round $r$.

**Step ❻: Gradient Compensate**. In this step, we utilize the $R_{EnD_i}^r$ of round $r$ to compensate for the lost gradient information due to **P₁**, **P₂**, and **P₃** operations. Specifically, in round $r+1$, we firstly receive the global update $\Delta \mathcal{G}_G^r$ of round $r$ and compute the compensate local update by $\Delta \mathcal{G}_{EnD_i}^{r+1} \leftarrow \beta R_{EnD_i}^r + (1-\beta) \Delta \mathcal{G}_G^r$, where $\beta$ is the momentum coefficient and satisfies $\beta \in [0,1)$. Then, $EnD_i$ can calculate the subsequent local model in round $r+1$ via the local update $\Delta \mathcal{G}_{EnD_i}^r$ in round $r$ by $W_{EnD_i}^{r+1} \leftarrow W_{EnD_i}^r - \eta \left( \Delta \mathcal{G}_{EnD_i}^{r+1} + \mathcal{G}_{EnD_i}^r \right)$.

Our concise pseudo code is presented in **Algorithm 1**.

*2) Edge Layer Steps*

Steps 1, 2, and 4 are the same as in the standard FL workflow in [4], step 5 is spired by [39], and step 3 is proposed in this paper. For easy understanding, we detail each step in the following.

**Algorithm 2 PA-iMFL in Edge Layer**

**Input:** training round $R$, mobile device client $EnD$, number of $EnD$ $m$, **P₁** privacy budget $\epsilon$, subsample ratio $\gamma$, interval rounds $t$, learning rate $\eta$, momentum coefficient $\beta$

**Initialize:** global model $W_G^0$

1 ############### Edge Layer Process ###############
2 **For** $r = 0,1,2,...,R$ **do**
3   # **STEP 1** : Gradient Compute and Copy
4   **If** $r == 0$ **then**:
5      $W_{EdD}^0 \leftarrow W_G^0$
6   **Else** round $r-1$:
7      Receive global update $\Delta \mathcal{G}_G^{r-1}$
8      $W_{EdD}^{r-1} \leftarrow W_{EdD}^{r-2} - \eta \left( \Delta \mathcal{G}_G^{r-1} + \mathcal{G}_G^{r-2} \right)$
9   **End if**
10   $\mathcal{G}_{EdD}^r \leftarrow \nabla l \left( W_{EdD}^{r-1} \right)$
11   $\Delta \mathcal{G}_{EdD}^r \leftarrow \mathcal{G}_{EdD}^r - \Delta \mathcal{G}_G^{r-1}$
12   Record $R_{EdD}^r \leftarrow \Delta \mathcal{G}_{EdD}^r$
13   # **STEP 2**: Gradient Update Aggregate
14   $\Delta \mathcal{G}_{EdD}^r \leftarrow \frac{1}{m+1} \left( \sum_{i=1}^{m} \Delta \mathcal{G}_{EnDi}^r + \Delta \mathcal{G}_{EdD}^r \right)$
15   # **STEP 3**: Gradient Transformation (**P₁, P₂, and P₃**)
16   $\Delta \mathcal{G}_{EdD}^r \leftarrow \left\{ \text{Lap}(\frac{\Delta f}{\epsilon}), PAS(\nabla W, \gamma), GSR(\nabla W, t) \right\}$
17   # **STEP 4**: Gradient Update Upload
18   Upload $\Delta \mathcal{G}_{EdD}^r$ to the *Cloud Server*
19   **If** $r = r+1$ **then**:
20      # **STEP 5**: Gradient Compensate.
21      $\Delta \mathcal{G}_{EdD}^{r+1} \leftarrow \beta R_{EdD}^r + (1-\beta) \Delta \mathcal{G}_G^r$, $\beta \in [0,1)$
22   **End if**
23 **End for**

**Step 1: Gradient Compute and Copy**. It is similar to **Step ❶** in end layer steps. In the round $r-1$, $EnD$ receives the whole global model $W_{EdD}^{r-1}$. After training $W_{EdD}^{r-1}$, $EdD$ obtain the new local update $\Delta \mathcal{G}_{EdD}^r$ for the training round $r$, and store the record $R_{EdD}^r \leftarrow \Delta \mathcal{G}_{EdD}^r$.

**Step 2: Gradient Update Aggregate**. $EdD$ aggregates the local update from $EnD_i$ and conducts FedAvg to compute the local update by $\Delta \mathcal{G}_{EdD}^r \leftarrow \frac{1}{m+1} \left( \sum_{i=1}^{m} \Delta \mathcal{G}_{EnDi}^r + \Delta \mathcal{G}_{EdD}^r \right)$.

**Step ③: Gradient Transformation (P₁, P₂, and P₃).** In this step, we integrate **Step ②, ③, ④** in the end layer, namely, $\Delta \mathcal{G}^r_{EdD} \leftarrow \left\{ \text{Lap}(\frac{\Delta f}{\epsilon}), PAS(\nabla W, \gamma), GSR(\nabla W, t) \right\}$.

**Step ④: Gradient Update Upload.** *EdD* uploads the local update $\Delta \mathcal{G}^r_{EdD}$ to the cloud layer. Up to now, the training process has finished in round $r$. Notably, *EdD* once receives the global update, *EdD* is going to broadcast to $EnD_i$.

**Step ⑤: Gradient Compensate.** It is similar to the **Step ⑤** in the end layer. In the round $r+1$, $\Delta \mathcal{G}^{r+1}_{EdD} \leftarrow \beta R^r_{EdD} + (1-\beta) \Delta \mathcal{G}^r_G$.

Our concise pseudo code is presented in **Algorithm 2**.

*B. Description of Privacy Components*

*1) P₂: Privacy Amplification Subsample (PAS)*

A composition of LDP with Laplace mechanism using additive minuscule perturbation and unbiased subsample results in the privacy amplification, where the capability of privacy-preserving increases exponentially instead of linearly [27], [31], [40], [41].

---

**Algorithm 3 P₂: PAS**

**Input:** subsample ratio $\gamma$, the raw local update $\Delta \mathcal{G}$

1  $d := \text{len}(\Delta \mathcal{G}) \in \mathbb{N}$
2  $k := \gamma \cdot d \in \mathbb{N}$  # define the subsample size
3  $\Delta \mathcal{G} := \langle g_1, g_2, ..., g_i \rangle \in [-clip, clip]$
4  **For** $i = 1, 2, ..., d$ **do**
5  $\quad \theta_i := \text{random}[0, \gamma]$  # Define the random coefficient
6  $\quad rs \leftarrow \theta \cdot (g_i)^2$  # Re-scaling
7  $\quad rss := [rs \text{ with descending order}]$
8  $\quad$ **If** $rs \geq rss[k]$: # Privatization
9  $\quad\quad p_i \leftarrow 1$
10 $\quad$ **Else**: $p_i \leftarrow 0$
11 $\quad$ **End if**
12 $\quad g_i \leftarrow B(1, p_i)$
13 **End for**
14 **Return** $\Delta \mathcal{G}$

---

With the intuition of the unidirectional sample method CRS-FL [23], we extend the bidirectional subsample method PAS. The detailed workflow of PAS is presented in **Algorithm 3**. Notably, *clip* is the $\ell_2$-norm clip value and we set constantly 1.0. $rs$ denotes the re-scaling value and the $rss$ is the set of $rs$. $p_i$ represents the subsample probability of $g_i$ and $B(\cdot)$ denotes the binomial distribution with probability $p_i$ in one dimension. Actually, the subsample method of PAS is the Poisson Sample (defined in **Definition 4.2**).

*2) P₃: Gradient Sign Reset (GSR)*

In communication-efficient unbiased sample methods [23], [26], the gradient signs are not compressed and going to disclose the true optimization directions of real gradients, which benefits for DRA adversaries [17]. Inspired by [33] in DP settings, we propose the GSR method for gradient sign transformation in a stochastic manner. Notably, different from [33], we do not conduct gradient quantization, and only change the sign. In addition, the total number of positive and negative gradients should be ensured in consistency.

The detailed workflow of GSR is shown in **Algorithm 4**.

---

**Algorithm 4 P₃: GSR**

**Input:** the raw local update $\Delta \mathcal{G}$, the length of the local update $d$, interval rounds $t$

1  $\Delta \mathcal{G} := \langle g_1, g_2, ..., g_i \rangle \in [-clip, clip]$
2  **For** $i = 1, 2, ..., d$ **do**
3  $\quad pos\_num \leftarrow count(g_i > 0)$  # The positive dimension numbers
4  $\quad neg\_num \leftarrow count(g_i < 0)$  # The negative dimension numbers
5  **End for**
6  **For** $i = 1, 2, ..., d$ **do**
7  $\quad \text{sign}(g_i) = \begin{cases} 1, & \text{with probability } \frac{pos\_num}{pos\_num + neg\_num} \\ -1, & \text{with probability } \frac{neg\_num}{pos\_num + neg\_num} \end{cases}$
8  $\quad g_i \leftarrow \text{sign}(g_i) \cdot |g_i|$
9  **End for**
10 **Return** $\Delta \mathcal{G}$

---

## V. EVALUATION

In this section, we first present the experimental settings. Then, evaluation results are demonstrated.

*A. Experimental Settings*

The following presents the experiment details including datasets, models, baselines, and hyperparameters.

*1) Datasets*

- **MNIST.** MNIST [43] dataset includes 28x28 handwritten digits with ten classes and has become the most well-known dataset in the classification task.
- **CIFAR-10.** CIFAR-10 [44] is made up of 10 classes of 32x32 images with three RGB channels and consists of 50,000 training samples and 10,000 testing samples.
- **CIFAR-100.** CIFAR-100 [44] has the same size of images. However, it exists 100 classes with 600 samples (500 training samples and 100 test samples) in each class. In other words, CIFAR-100 is a more sophisticated dataset than CIFAR-10 and thus evaluates the FL system more fairly.
- **ImageNet.** ImageNet [46] is a large-scale dataset in computer vision. It includes more than 14 million images. In 2010, ImageNet was a benchmark of the famous algorithm competition ImageNet Large Scale Visual Recognition Challenge (ILSVRC).

The datasets we utilized are the famous visual benchmarks in computer vision and deep learning. It will increase extremely the visual experience more intuitively, especially in DRA analysis. In addition, several SOTA DRAs and privacy-preserving schemes against DRA choose those datasets for evaluating attack or defense performance [14], [16]–[18].

Notably, non-independent identical distribution (non-i.i.d) data is a key feature of the realistic FL scene. So we adopt a non-i.i.d split of training data in the PA-iMFL training process.

*2) Baselines*

The baselines to be compared are divided into three categories: a) Data reconstruction attack baselines, b) Privacy-preserving FL against DRA baselines, and c) Communication-efficient FL baselines. Details are as follows.

*a) Data reconstruction attack baselines*

**IG** (*NeurIPS'23*, [16]). IG combines the cosine similarity of gradients and Adam optimizer to minimize the divergence between the target image and dummy data. IG achieves a huge improvement compared to the genetic DRA method [15] and has become the famous baseline in the past years.

**GLAUS** (*TKDE'23*, [17]). GLAUS is proposed to conduct DRA in the unbiased sample communication efficiency method, such as [23] and [26]. It can narrow the range of dummy gradients from aggregating global updates and deduce the gradient sign due to the unchangeability of the sample gradient. From those observations, GLAUS reconstructs the private data in a sample ratio of 5%.

**ROG** (*USENIX Security'23*, [18]). ROG attack, a new DRA framework, is proposed by optimizing the dummy data using low-dimensional representation. Specifically, the dummy data is encoded by a lossy bicubic downsampling function and then mapped into the original size of the target private data.

*b) Privacy-preserving FL against DRA baselines*

**ATS** (*TPAMI'23,* [14]). ATS is a novel defense method to mitigate DRA. It draws support from data augmentation policies to transform original images into other presentations. For example, they optimize the maximal divergence between target transformation data and dummy data reconstructed by SOTA DRA and then output the transformation policy (such as rotate and deviate) for other original images. In addition, ATS proposes an automatically discover-qualified policy from more than 50 kinds of data augmentation policy according to two well-designed new metrics. Notably, unlike other defense methods, ATS modifies the private data and consumes more computation overhead in FL participants.

*c) Communication-Efficient FL baselines*

**FedAVG** [4]. Benefiting from the simplicity of averaging the weights of participants, FedAVG is the most used baseline without privacy and communication efficiency.

**CRS-FL** [23]. CRS-FL conducts unbiased Poisson sample to reduce the transmitting bytes with little accuracy decline. Notably, it only achieves unidirectional sample. In other words, the resource-constrained participants transmit their local update using CRS but the server broadcasts the whole global update in return.

*3) Supplementary settings on DRA*

**Label**. Label inferring is a crucial task in DRA [15], [17]. In this experimental evaluation on DRAs, we assume that each SOTA DRA method can obtain the true labels of target private data. In this assumption, the attack capability of DRAs, including IG, GLAUS, and ROG, is improved heavily.

**Hyper-parameters**. In our evaluation, we use the official hyper-parameters of each SOTA DRA method. In addition, we exclusively choose the official code for the corresponding DRA method. It will reduce the potential errors caused by experimental replication and then affect the fairness of experimental comparisons. For example, IG[1], GLAUS[2], ROG[3], and ATS[4].

**Dataset**. To improve the convincingness of our PA-iMFL method, we choose the main dataset mentioned by DRA or privacy methods in their paper and code. For example, MNIST in GLAUS, CIFAR-10/100 in IG and ATS, and ImageNet in ROG (we select the StanfordCars to relate our iMFL scene).

Notably, due to the various DRA aiming at different purposes, for example, the communication-efficient method with a high sample ratio in GLAUS, general DRA in ROG, providing the privacy-preserving using data augmentation in ATS, we place those three as the same important competitors.

In one word, we will divide three subsections to evaluate the privacy-preserving performance of PA-iMFL in Section V.C. In one subsection, we will mainly utilize the represented dataset of one competitor to accelerate the statement of our paper.

*4) Supplementary Settings on Communication Efficiency*

In communication efficiency comparisons, we adopt the same hyper-parameters with CRS-FL [23] and the only difference is that we use a subsample ratio $\gamma$ to replace the sample size. The detailed information is stated in Table I. Notably, the CRS-FL uses a unidirectional sample with an LDP mechanism, while PA-iMFL utilizes a bidirectional subsample. Therefore, PA-iMFL will have less communication overhead than SOTA CRS-FL theoretically. In other words, the upper bound of CRS-FL is $O(d+d\gamma)$, while the lower bound is $\Omega(d+0)$ with $p_\Omega = \left(1/\left(1-e^{-\epsilon}\right)\right)^{d\gamma}$ [23]. However, the upper bound of PA-iMFL is $O(d\gamma + d\gamma)$ while the lower bound is $\Omega(0+0)$ with $p_\Omega = \left(1/\left(1-e^{-\epsilon}\right)\right)^{2d\gamma}$, which is extremely low probability.

TABLE I    THE HYPER-PARAMETERS IN DRA

| Dataset | Model | Data Split | Subsample Ratio $\gamma$ | Other Hyper-parameters |
|---|---|---|---|---|
| CIFAR10 | ResNet 9 [23] | Non-i.i.d | 0.07, 0.10, 0.15, 0.20, 0.25 | Epoch 24, Iteration total 2400, Participants 200, Learning Rate 0.01, with decay 1e−4, Local batch size 50, Aggregation FedAVG, $\epsilon = 0.1$ $\gamma \in (0,1]$, $t = 100$. |

*B. Evaluation Metrics*

MSE (Mean Square Error) is used for measuring the distance between two vectors. In our image evaluation task, as the statement in Eq. (8), $s_1$ and $s_2$ denotes the number of horizontal pixels and vertical pixels in an image, respectively. Besides, $x(i, j)$ denotes the original input while $x'(i, j)$ is the dummy image reconstructed by the adversary $\mathcal{A}$. Notably, in [17], they utilize MSE as the major metric for DRA.

---

[1] IG code: https://github.com/JonasGeiping/invertinggradients.
[2] GLAUS code: https://github.com/Echotoken/GLAUS.
[3] ROG code: https://github.com/KAI-YUE/rog.
[4] ATS code: https://github.com/gaow0007/ATSPrivacy.

$$MSE = \frac{1}{s_1 s_2} \sum_{i=1}^{s_1} \sum_{j=1}^{s_2} [x(i,j) - x'(i,j)]^2 \quad (8)$$

*C. Comparisons on Privacy-preserving (against DRA)*

In this section, we present the evaluation results of privacy-preserving performance (**Goal 1**) against three SOTA DRAs, including GLAUS, IG, and ROG.

*1) GLAUS in MNIST*

We first conduct the MMS against GLAUS with $\gamma \in (0,1]$. As presented in GLAUS [17], the unbiased sample method MMS[26] probably causes privacy leakage with the subsample ratio $\gamma = 0.05$. In addition, we also perform the GLAUS when $\gamma = 0.01$. GLAUS will reconstruct the private image of MNIST. The related results are illustrated in **Fig.5**. However, if the sample ratio $\gamma$ increases, the GLAUS loses effectiveness. In realistic scenes, adversaries always cannot obtain the precise sample ratio.

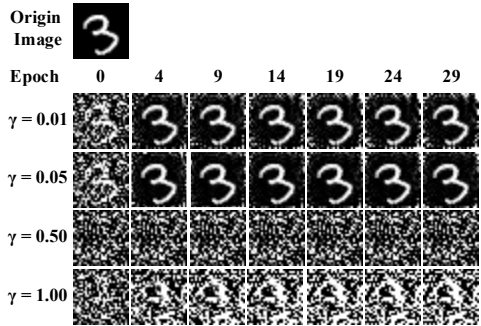
**Fig.5**. The GLAUS attack against MMS in MNIST with different $\gamma$.

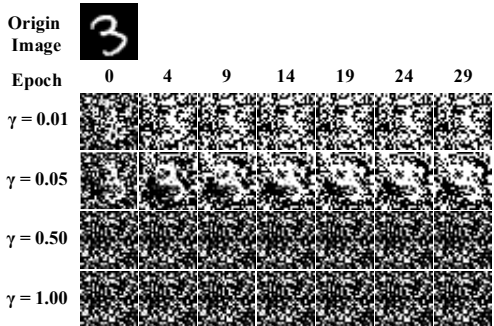
**Fig.6**. The GLAUS attack against PA-iMFL (ours) in MNIST with different $\gamma$.

In **Fig.6**, we conduct experiments utilizing PA-iMFL against GLAUS in MNIST. Obviously, our PA-iMFL performs better privacy-preserving ability than MMS. On the condition of different subsample ratios $\gamma$, the adversary with GLAUS attack capability cannot reconstruct the private data.

*2) IG in CIFAR100*

In this subsection, we evaluate the privacy-preserving performance against IG, which is the main comparator in ATS [14]. Note that ATS explores the best data transformation method based on different data augmentation policies, to mitigate the DRA. We implement the open-source code of ATS and search for the best transformation policies in CIFAR100 against IG, namely 3-1-7. The examples, after data transformation, are illustrated in **Fig.7**.

In **Fig.8**, we summarize the MSE of various defense models against IG. For example, we give the IG *v.s.* FedAVG (no defense) in the second row. The remaining ones IG *v.s.* ATS (policy 3-1-7), and IG *v.s.* PA-iMFL ($\gamma = 0.07$ and $\epsilon = 0.1$). Notably, the higher MSE value means more divergence between reconstructed data and the origin image. Moreover, the MSE is calculated by randomly selecting 50 reconstructed images.

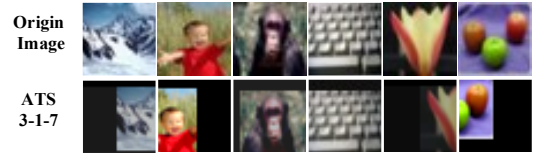
**Fig.7**. The examples of policy 3-1-7 and 43-18-18 in ATS.

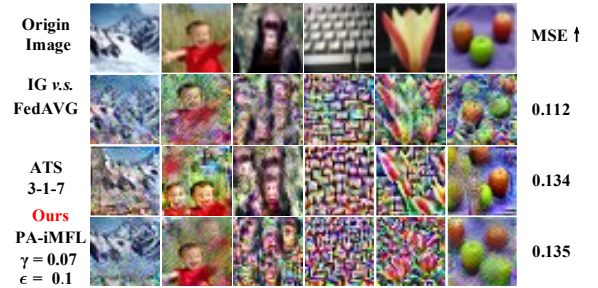
**Fig.8**. MSE results of different defense models against IG attack in CIFAR100.

Thus, our PA-iMFL obtains the better privacy-preserving performance with MSE 0.135, which is the same defense level with SOTA defense ATS. However, ATS needs to search for the best data augmentation policy from more than 50 policies, which is time-consuming and is unlikely achieved in resource-constraint participants.

*3) ROG in ImageNet and StanfordCars*

Against the other SOTA DRA method ROG, we evaluate dataset ImageNet and StanfordCars.

Similarly, we give the main results in **Fig.9**. The first row presents the origin images of the target dataset, and the following rows are the ROG *v.s.* FedAVG (no defense), ROG *v.s.* PA-iMFL ($\gamma = 0.07$ and $\epsilon = 0.1$). Moreover, in ImageNet and Stanford, the MSE is calculated by randomly selecting 16 reconstructed images.

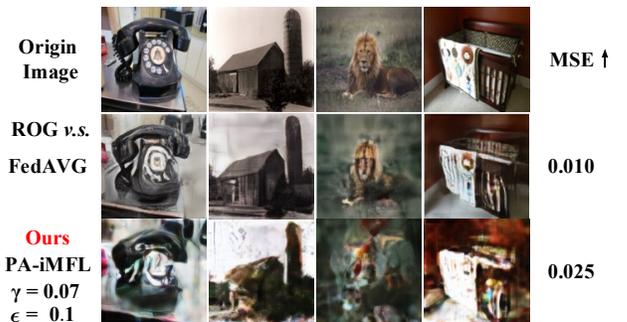
**Fig.9**. MSE results of different defense models against ROG in ImageNet.

*D. Comparisons on Communication Overhead*

In this section, to demonstrate the satisfaction of **Goal 2** and **Goal 3**, we conduct the communication overhead and accuracy analysis. As demonstrated in Section V.A.5), we maintain the same hyper-parameters with CRS-FL [23].

Firstly, we present the accuracy and transmitting bytes results in **Fig.10**. Under the subsample ratio $\gamma = 0.07$, the accuracy of PA-iMFL reaches the same level after 1,700 iterations. However, there exists a gap from the FedAVG (we set FedAVG as the convergence baseline). Specifically, the accuracy of FedAVG is 84.95%, while CRS-FL is 82.55% and 81.93% of PA-iMFL.

In communication overhead, the CRS-FL transmits over 1,754M bytes in a total of 2,400 iterations while PA-iMFL transmits 457M bytes, which improves the communication efficiency by 3.8×.

Then, we evaluate the accuracy and communication overhead in various $\gamma$, where $\gamma = \{0.07, 0.10, 0.15, 0.20, 0.25\}$. Our PA-iMFL obtains the convergence when $\gamma \geq 0.15$. The accuracy divergence between CRS-FL and PA-iMFL is quite small in **Fig.11.(a)**, while the communication overhead is an extremely significant gap. As illustrated in **Fig.11.(b)**, the communication overhead of PA-iMFL improves 3.8×, 3.3×, 2.8×, 2.4×, and 2.1×, separately, than CRS-FL, and remarkably lower than FedAVG baseline of more than 6,043M bytes. Here, the **Goal 2** and **Goal 3** of our PA-iMFL have been achieved.

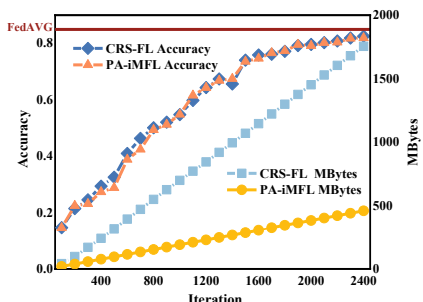

**Fig.10**. The accuracy and communication overhead result under $\gamma = 0.07$.

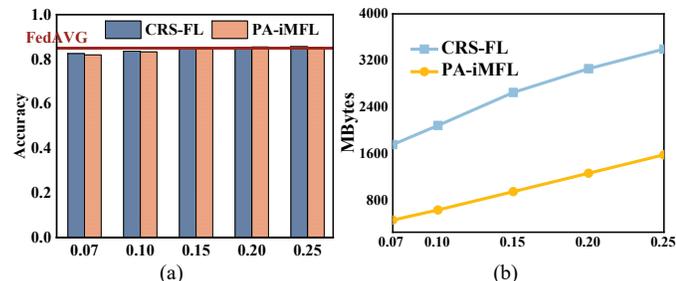

**Fig.11.** **(a)** The accuracy results under different $\gamma$. **(b)** The transmission bytes under different $\gamma$.

## VI. RELATED WORK

This section focuses on related works about DRA and the corresponding defense mechanism. In addition, we present the existing studies of communication-efficient methods on sample and privacy amplification subsample.

*A. DRA*

Data Reconstruction Attack (DRA) discovered in deep learning models [15] is also named Deep Leakage from Gradient (DLG) or Gradient Leakage Attack (GLA) in some studies. It allows the adversary to reconstruct the training data by stealing the training gradients. In the FL scene, DRA is more common because of the transmitting demand between participants and the central server in FL. The existing studies of DRA can be classified into three paradigms: optimization-based, analytics-based, and generation-based[36].

In optimization-based DRA, adversaries firstly stochastically create dummy data and the corresponding dummy label and then use several optimizers (such as L-BFGS [15] and Adam [16]) and/or distance policy (cosine similarity [16]) to shorten the divergence between the dummy data and the true data. Moreover, this kind of attack highly depends on the correctness of the dummy label. Hence, there exist studies focusing on label inferring from gradients or compression gradients [17], [19] instead of optimizing directly [15].

In analytics-based DRA, Phong *et al*. [49] provide the pioneering work. They analyze and calculate the original input from gradients of the specific layer, which is accurate and fast. However, those analytics-based DRAs only focus on simple and linear models [36].

Recently, facing DRA defense methods, such as high compression defense models, some generation-based methods were proposed. The optimization of the generation-based method is to minimize the divergence between dummy gradients and the real compressed gradient, such as HCGLA [19] and GLA [36]. The main difference between generation-based DRA and common DRA is that generation-based DRA utilizes the compressed gradient, for example, Generative Adversarial Network (GAN), to train a generation model for data reconstruction.

*B. Defense Against DRA*

As demonstrated by Zhu *et al.* [15] and Liu *et al.* [24], cryptology, additive noise, and gradient compression perform a significant defense performance.

Cryptology-based defense methods are the most secure strategy for protecting from information leakage [15]. However, they rely on abundant computation resources because of the high computation overhead of homomorphic encryption techniques [24], [50], [51] and highly customized [15].

Additive noise and gradient compression are the common strategies for DRA defense. Specifically, weak DP [13], [48] adds the perturbation directly without considering the privacy budget, while DP [52], [53] adds noise or utilizes randomization mechanisms in participants to prevent DRA. Gradient compression-based methods include prune [25] and top-k sparsification [54]. However, additive noise and gradient compression always sacrifice the model accuracy against DRA. It is a crucial problem to tradeoff the model accuracy and privacy-preserving capability.

Except above-mentioned methods, some researchers focus on data augmentation-based methods for protecting from DRA [14] in FL. In other words, the participants train the private data after data augmentation. It means that the training gradients only retain little input information to prevent privacy leakage.

However, it is time-consuming in policy searching and data transformation.

*C. Communication-efficiency in FL*

There exist communication bottlenecks in the resource-constraint FL system. Sample [23], [26], [27], or subsample [55] have been adopted to reduce the communication overhead between the central server and participants.

In the sample-based compression method, Zhao *et al.* [26] propose a MinMax unbiased sample in FL to speed up the uplink from participants to the parameter server without compromising the model accuracy. Then, Wang *et al.* [23] provide an advanced sample method CRS-FL based on MinMax using conditional random probability to lower the communication overhead further. In addition, CRS-FL provides the LDP guarantee but does not give the privacy experiments against DRA. Chen *et al.* [27] propose a Poisson Sampling method to reduce the communication overhead and give the Rényi DP guarantee conditions of their mechanism.

Subsample is proposed to amplify the privacy protection capability [31], [32]. It usually combines some basic DP mechanisms, such as DP with Laplace/Gaussian, random response, and shuffle, to gain greater privacy-preserving ability while consuming less privacy budget. Phuong et al. [56] utilize a subsample policy to decrease the communication overhead and protect the private dataset in distributed learning. However, they cannot provide an adequate privacy condition guarantee and enough experiments to reveal the efficiency and privacy results.

## VII. CONCLUSION

This paper studies the defense approach against Data Reconstruction Attack (DRA) in the iMFL scene, where the resource-constrained edge device owns private data and joins the training process. We propose a Privacy Amplification scheme on iMFL (PA-iMFL) including three privacy components, local differential privacy with Laplace mechanism, privacy amplification subsample, and gradient sign reset. Extensive results demonstrate that against State-Of-The-Art (SOTA) DRAs, PA-iMFL can effectively protect from private data leakage and reach the same protection capability as the SOTA defense model. Moreover, benefitting from privacy operations, namely privacy amplification subsample in edge devices, PA-iMFL promotes $3.8\times$ communication efficiency than the SOTA method with a subsample ratio of 0.07 and $2.8\times$ communication efficiency without compromising model accuracy.